\documentclass[%
groupedaddress,
 amsmath,amssymb,
 aps,
pre,
12pt
]{revtex4-1}

\usepackage[utf8x]{inputenc}
\usepackage{enumerate}
\usepackage{amsmath}
\usepackage{amssymb}
\usepackage{graphicx}
\usepackage{amsthm}

\newcommand{\dx}{\Delta x}
\newcommand{\dt}{\Delta t}

\newcommand{\ensav}[2]{\left\langle #1 \right\rangle_{#2}}

\newcommand{\dsum}{\displaystyle\sum}

\newcommand{\av}[1]{\left< #1 \right>}

\begin{document}

\title{Multinomial Diffusion Equation}

\author{Ariel Balter}
   \affiliation{Pacific Northwest National Laboratory P.O. Box 999, Richland, WA 99352}
   \email{ariel.balter@pnl.gov}
\author{Aleaxndre Tartakovsky}
   \affiliation{Pacific Northwest National Laboratory P.O. Box 999, Richland, WA 99352}
   \email{alexandre.tartakovsky@pnl.gov}



\begin{abstract}
We describe a new, microscopic model for diffusion that captures diffusion induced fluctuations at scales where the concept of concentration gives way to discrete particles.  We show that in the limit as the number of particles $N \to \infty$, our model is equivalent to the classical stochastic diffusion equation (SDE).  We test our new model and the SDE against Langevin dynamics in numerical simulations, and show that our model successfully reproduces the correct ensemble statistics, while the classical model fails.
\end{abstract}

\maketitle
\section{Introduction}
Fluctuations in concentration become important when modeling systems with small particle density, due either to small concentration or small spatial scale.  They can have a significant effect on the average behavior of many diffusion-reaction systems.  Fluctuations in chemical reaction kinetics result from both the intrinsic stochastic nature of chemical reactions and concentration fluctuations.  Concentrations fluctuations, in turn, result from both reaction fluctuations and the the random, thermal motion of reaction species.  Thus, reaction-diffusion systems couple kinetics and diffusion at both the deterministic (macroscopic) and stochastic (microscopic) scales.  When nonlinear reactions are present, initial fluctuations can induce instabilities that lead to interesting macroscopic behavior such as pattern formation and oscillations \cite{Dewit2007}.  When diffusion induced fluctuations are also present, we may see \emph{different} behavior than deterministic models of the same systems predict \cite{Awazu2010, Gopich2001, Togashi2001, Hecht2010}.  This highlights the importance of having good models for diffusion induced fluctuations.

Various Lagrangian- and Eulerian-frame representations exist for theoretical and numerical modeling of reaction-diffusion systems at the fluctuation scale.  One fundamental way to model reaction and diffusion is through Langevin dynamics, i.e. particle tracking.  This is a Lagrangian-frame representation that tracks the motion of individual particles whose dynamics is described by the Langevin equation (possibly in over-damped form), and models reactions based on some probabilistic or deterministic function of inter-particle distance.

In many situations it is more convenient to work in an Eulerian frame where one is interested in the concentration of material at a point in space (or the number of particle in a small volume).  The classical Eulerian description of diffusion is the diffusion PDE which one can derive by considering an ensemble of particles in Brownian motion.  However, this is a macroscopic model for average particle density, and does not include fluctuations.  A mesoscopic description that includes fluctuations is the Multivariate Master Equation (MME), a spatially discrete (Eularian) continuous time Markov chain.  The MME has been used to obtain some important rigorous results concerning the onset of instabilities in reaction-diffusion systems \cite{VanKampen2007}. The advent of powerful computers has enabled numerical sampling of master equations to become feasible.  Exact sampling methods, such as stochastic simulation algorithms (SSAs) exist, but are usually slow and, more importantly, progress in random time steps \cite{Gillespie2007}.  This complicates multi-scale modeling, especially where one would like to couple microscopic to mesoscopic to macroscopic models where the transitions between mesoscopic and macroscopic regimes may change dynamically with space and time.

One can derive a stochastic diffusion equation (SDE) from the MME as a thermodynamic limit.  The SDE adds a stochastic flux to the classical diffusion equation.  The SDE can be discretized and use in numerical simulations where one wished to model diffusion induced fluctuations.  It also evolves in fixed time steps.  Therefore, the SDE can seamlessly integrate with a forward-Euler finite-difference integration of the deterministic diffusion PDE.  The SDE is little more expensive than for deterministic diffusion -- generating Gaussian random variables being the additional expense.  However, since the SDE represents the thermodynamic limit, a valid theoretical question is to what degree models very small particle densities.

We have found a new representation, called the multinomial diffusion equation, (MDE) that describes the evolution of the numbers of particles in a spatially discretized field in fixed time steps.  Using numerical simulations, we compare the diffusion induced fluctuations in both the MDE and SDE to those observed in a particle tracking model.  We find that our new MDE more closely reproduces diffusion induced fluctuations than the SDE.  Therefore, we conclude that the MDE provides a theoretical middle ground between the MME and the SDE.  We also found that the MDE can be used as an efficient and accurate finite difference method for modeling diffusion at the particle scale.  It is comparable to a particle simulation in accuracy, yet is almost as computationally efficient as the SDE (in a finite-difference discretization), and also evolves (synchronously) in fixed time steps.

For the remainder of this article, we will continue to use the term ''diffusion'' rather than ''Brownian motion'', which might more accurately specify that we are looking at the diffusion of \emph{particles}, as opposed to \emph{heat}, for instance. However, we will from now on use the term ``particle density`` instead of ``concentration`` to emphasize that we are in the regime of individual particles.

\section{Multivariate Master Equation}
The Multivariate Master Equation (MME) models the numbers of particles in $M$ voxels of size $\dx$ in a spatially discretized domain of size $L=M\dx$ \cite{Gardiner1996}.  The state of the system is a spatial field of particle numbers recorded in the vector $\vec N = [N_1, N_2, ..., N_M]$ where $N_i$ is the number of particles in the $i^\text{th}$ voxel (centered at $\dx(i + \frac{1}{2})$).  In a transition event, a \emph{single} particle hops from the $j^\text{th}$ voxel to the $i^\text{th}$ voxel.  Such a transition changes the state from $\vec N$ to $[N_1, N_2, ..., N_j-1, ..., N_i + 1, ..., N_M]$.  More compactly, $\vec N \to \vec N + \vec \Delta$, where $\vec \Delta$ has only two nonzero elements: $\Delta_j = -1$ and $\Delta_i = 1$.  Let $\mathcal{P}[\vec \Delta | \vec N(t),\dt]$ be the probability that that the transition $\vec N(t) \to \vec N(t) + \vec \Delta$ occurs during the next small time increment $\dt$.  We would like to have an expression for $P(\vec N(t+\dt),\vec N(t)) = P[\vec N(t+\dt)|N\vec (t)]\,P(\vec N(t))$. Since $\vec N(t+\dt) = \vec N(t) + \vec \Delta$, $P(\vec N(t+\dt) | \vec N(t)) \equiv P(\vec \Delta | \vec N(t+\dt))$.  Therefore, we have

\begin{align}
 \label{eqn:MME_total}
 & P(\vec N(t+\dt),t+\dt | \vec N(t)) & \\ \nonumber
 & = \dsum_{\vec \Delta} \mathcal{P}[\vec \Delta | \vec N] P(\vec N(t)) \\ \nonumber
 & = \dsum_i \dsum_j \mathcal{P}[i \to j | \vec N] P(\vec N(t)) \\ \nonumber
\end{align}

\noindent where $i \to j$ stands for $\Delta_i = -1, \Delta_j = 1$.  As in pure Brownian motion, we will assume that individual particles do not interact.  With this assumption, $\mathcal{P}[i \to j]$ depends only on $N_i(t)$.  Also, when the linear dimension of a voxel is larger than the mean free path of a Brownian particle, we need only include nearest neighbor hops, i.e. $|i-j| \in {0,1}$ .  The mean free path ($\lambda$) for Brownian motion is a measure of the distance a particle can travel after an impulse from the surrounding fluid.  A good estimate assumes the particle starts at thermal speed $\sqrt{\frac{3 k_b T}{m}}$, giving $\lambda \sim \frac{\sqrt{3 k_b T}}{6 pi \eta a}$ where $\eta$ is the fluid viscosity, and $a$ is the particle diameter. A "particle" must be larger than molecular size $\sim 10^{-10}m$. Using molecular size, and the density of stone ($\sim 10^6 k/ m^{-3}$), the mean free path is $\sim 10^{-10}m$ -- no more than the diameter of the particle itself!  With these assumptions, and using $\mathcal{P}(i \to i) = 1 - \mathcal{P}[i \to i-1] - \mathcal{P}[i \to i+1]$ we transform Eq. \eqref{eqn:MME_total} to

\begin{align}
 \label{eqn:MME_CK}
& P(\vec N(t + \dt) ,\vec N(t)) = \dsum_i \\ \nonumber
& \mathcal{P}(i-1 \to i | N_{i-1}(t)) P(N_{i-1}(t)) \\ \nonumber
& + \mathcal{P}(i+1 \to i | N_{i+1}(t)) P(N_{i+1}(t)) \\ \nonumber
& + (1 - \mathcal{P}(i\to i-1 | N_i(t)) -  \mathcal{P}(i\to i-1 | N_i(t)P(N_i(t)) \\ \nonumber
\end{align}

\noindent which has the form of a Chapman-Kolmogorov equation for the interval from $t$ to $t + \dt$.

An informal way to define a transition \emph{rate} $\mathcal{W}[j \to i]$ from a transition probability $\mathcal{P}[j \to i]$ is to say that $\mathcal{W}[j \to i] = \frac{d}{dt}\mathcal{P}[j \to i]$.  One can make this rigorous when $\mathcal{P}[j \to i] \approx q_{j,i}\dt + \mathit{o}(\dt)$ and $\mathcal{P}[i \to i] \approx 1 - q_{i,i}\dt + \mathit{o}(\dt)$.  In this manner, Eq. \eqref{eqn:MME_CK} gives

\begin{align}
 \label{eqn:MME_local}
 \frac{P(\vec N(t),t)}{dt} & = \dsum_j  \\ \nonumber
 & ~~~ \mathcal{W}[j+1 \to j| N_{j+1}(t)] P(N_{j+1}(t),t)\\ \nonumber
 & - \mathcal{W}[j \to j+1| N_j(t)] P(N_j(t),t)\\ \nonumber
 & - \mathcal{W}[j \to j-1| N_j(t)] P(N_j(t),t)\\ \nonumber
 & +  \mathcal{W}[j-1 \to j| N_{j-1}(t)] P(N_{j-1}(t),t)\\ \nonumber
\end{align}

\noindent which is known as the multivariate master equation for diffusion.

The theory of continuous time Markov chains allows us to decompose this process into two independent random processes: (1) a random waiting time until the next transition and (2) a random selection of which transition occurs \cite{Breiman1969}.  The distributions for these random numbers depend on the transition probabilities $\mathcal{W}(\vec \Delta | \vec N)$.  This is the basis of exact sampling algorithms such as the Gillespie algorithm \cite{Gillespie2007}.

\section{Multinomial Diffusion Equation}
We now describe a representation that is spatially discrete (as is the MME), but evolves in fixed time steps (as does particle tracking).  We use the same definitions as we did in deriving the MME, except that we do not restrict to single particle exchanges.  Instead, $\Delta = (\Delta_1, \Delta_2, \ldots, \Delta_M)$ where $\Delta_i$ can have any value $0\le \Delta_i \le N$, so long as $\sum_i \Delta_i = 0$.  We again invoke the assumption that when $\dx$ is larger than the mean free path, only consider nearest neighbor exchanges.  Since we are now working with fixed time steps, we will use time steps as an index placed as a superscript.  Let the vectors $\vec L^t$ and $\vec R^t$ record the random number of particles that jump out of voxel $i$ to the left and right respectively (with suitable boundary conditions) in the time interval from $t$ to $t + \dt$.

Let $\kappa \dt$ be the probability that an individual particle can jump into the next voxel during an interval of size $\dt$.  In this case, the probability that $L_i^t$ particles jump to the left out of voxel $i$, \emph{and} $R_i^t$ to the right is given by the multinomial multinomial distribution $\mathcal{M}(N_i^t,k\dt,k\dt)$:

\begin{equation}
 \label{eqn:multinomial}
 P[L_i^t, R_i^t] = N_i^t ! 
\frac{(k\dt)^{L_i^t} \, (k\dt)^{R^t_i} \, (1 - k\dt)^{N_i^t - L^t_i - R^t_i}}{L^t_i ! \, R^t_i ! \, (N_i^t - L^t_i - R^t_i)!}
\end{equation}

\noindent This is the essential feature of the MDE.

Can we derive a master equation for the MDE?  Let us define the shift operators $\mathbb{L}$ and $\mathbb{R}$ such that 

\begin{subequations}
 \label{eqn:shift_operators}
\begin{equation}
 [\mathbb{L}\vec L]_i = [\vec L]_{i+1}
\end{equation}
\begin{equation}
 [\mathbb{R}\vec L]_i = [\vec L]_{i-1}
\end{equation}
\begin{equation}
 [\mathbb{L}\vec R]_i = [\vec R]_{i+1}
\end{equation}
\begin{equation}
 [\mathbb{R}\vec R]_i = [\vec R]_{i-1}
\end{equation}
\end{subequations}

\noindent and $\mathbb{L}\mathbb{R} = \mathbb{R}\mathbb{L} = \mathbb{I}$.  For example, $\mathbb{L}$ pulls the value of $\vec L$ in slot $i+1$ back to slot $i$, and likewise for $\vec R$.  This gives the relationship $\vec \Delta = \mathbb{L}\vec L^t + \mathbb{R} R^t - (L^t + R^t)$.  Conditioning on $\vec L$ and $\vec R$, and again using only nearest neighbor exchanges, we can write

\begin{multline}
 \label{eqn:MDE_CK}
 P(\vec N(t+\dt),t+\dt | \vec N(t),t)  = \dsum_i \\ 
 \mathcal{P}[L_{i-1}^t, R_{i-1}^t | N_{i-1}^t] P(N_{i-1}^t,t) \\ 
+ \mathcal{P}[L_{i+1}^t, R_{i+1}^t| N_{i+1}^t] P(N_{i+1},t) \\ 
+ (1 - \mathcal{P}[L_{i}^t, R_{i}^t | N_i^t) )P(N_i^t,t) \\ 
\end{multline} 

\noindent This also has the form of a Chapman-Kolmogorov equation, and is comparable to Eq. \eqref{eqn:MME_CK}. However, from Eq. \eqref{eqn:multinomial} we see that the transition probabilities in Eq. \eqref{eqn:MDE_CK} are \emph{not} $\mathit{o}(\dt)$.  Therefore, we can not construct a master equation as we did for the MME. 

Nevertheless, we still \emph{can} generate exact realizations of $\vec N(t)$.  At each time step, we can generate the random vectors $\vec L^t(t)$ and $\vec R^t(t)$ according to Eq. \eqref{eqn:multinomial}, and then perform the updates

\begin{equation}
 \label{eqn:vector_update}
 \vec N^{t+\dt}  = \vec N^t + \mathbb{L} \vec L^t + \mathbb{R} \vec R^t - (\vec L^t + \vec R^t)
\end{equation}

\noindent We will also write this term-by-term

\begin{align}
 \label{eqn:MDE_update}
 N_i^{t + \dt}  & = \\ \nonumber
& N_i^{t} + L^t_{i+1} - R^t_{i,t} - L^t_{i,t} + R^t_{i-1}
\end{align}

\section{Stochastic Diffusion PDE}
There is a classical stochastic diffusion PDE that can be derived by various methods in the thermodynamic limit of $N \to \infty$.

\begin{equation}
\label{eqn:SDE}
 \frac{\partial \rho(x,t)}{\partial t} =  D\frac{\partial^2}{\partial x^2} \rho(x,t) + \frac{\partial}{\partial x} \sqrt{2 D \rho(x,t)}  \, \xi (x,t)
\end{equation}

\noindent Keizer derives Eq. \eqref{eqn:SDE} using thermodynamic potentials \cite{Keizer1987}.  Gardiner derives Eq. \eqref{eqn:SDE} from MME using a Van Kampen system size expansion \cite{Gardiner1996}.  Ironically, these derivations employ the limit $N \to \infty$, even though fluctuations are only significant when $N \ll \infty$.  This suggests there is a lower limit of particle density where this description will apply.  For instance, we wonder if Eq. \eqref{eqn:SDE} can accurately model diffusion induced fluctuations as well as a particle tracking model when the particle density is very small.

%
%


\section{Thermodynamic Limit}
It is common to use a multidimensional Gaussian distribution to approximate a multinomial distribution \footnote{Just as a single Gaussian approximates a binomial.}.  A multinomial distribution $P[n_1,n_2,\ldots,n_M] = N! \, \prod_{i=1}^M \frac{p_i^{n_i}}{n_i !}$, with $\sum_{i=1}^M n_i = N$, can be approximated by a multivariate Gaussian with means $\mu_i = N p_i$, variances $\Sigma_{i,i} = N p_i(1-p_i)$, and covariances $\Sigma_{i,j} = -N p_i p_j$.  The approximation becomes better as $N$ gets larger, but worse as each $p_i$ gets smaller.

Let us consider what happens if we make this approximation in Eq. \eqref{eqn:MDE_update}.  Since $k$ is a probability rate, as $t \to 0$, $k\dt$ becomes very small.  In this limit, $\Sigma_{i,i} \approx N k\dt + \mathit{o}(\dt)$, and $\Sigma_{i,j} \approx \mathit{o}(\dt^2)$.  The multinomial random variables in Eq. \eqref{eqn:MDE_update} become independent Gaussian random variables, and we have


\begin{align}\label{eqn:norm_approx_2}
 N_i^{t+\dt} & = N_i^t + \kappa\dt[ N_{i+1}^t - 2N_i^t + N_{i-1}^t] \\ \nonumber
& +  \sqrt{N_{i+1}^{} \kappa \dt} \, \xi _{i+1}^t \\ \nonumber
& -  \sqrt{N_{i}^{t} \kappa \dt} \, \xi _{i}^{t} \\ \nonumber
& -  \sqrt{N_{i}^{t} \kappa \dt} \, \xi _{i}^{t} \\ \nonumber
& +  \sqrt{N_{i-1}^{t} \kappa \dt} \, \xi _{i-1}^t \\ \nonumber
\end{align}

\noindent From the deterministic part of Eq. \eqref{eqn:norm_approx_2} we learn that $k\dt \equiv \frac{D\dt}{\dx^2}$.  This is expected, since a particle has a high probability of traveling a distance $\dx$ in a time interval $\dt = \sqrt{D/\dx^2}$.

Conservation of mass requires that we cannot remove more than $N_i$ particles from voxel $i$ in any time step.  In the MDE, using the multinomial distribution \emph{ensures} that mass is conserved.  We ask how large must $N$ be so that Eq. \eqref{eqn:norm_approx_2} will \emph{almost never} violate conservation of mass?  Conservation of mass for the deterministic part of Eq. (\ref{eqn:norm_approx_2}) requires
\footnote{Interestingly, we see that the Courant condition \cite{press_numerical_1992} for the stability of numerical discretization equivalent to the deterministic part of the Gaussian version corresponds to conservation of mass in the model.} 

\begin{equation}
 \label{eqn:courant}
 \frac{D \dt}{\dx^2}2N < N \implies \frac{D \dt}{\dx^2} < 1/2
\end{equation}

\noindent However, due to the fluctuating part of Eq. \eqref{eqn:norm_approx_2} there is a finite probability that $N_i^{t+\dt}<0$, even if Eq. \eqref{eqn:courant} holds.  To obtain a rough estimate for how rare such an even would be, we require that the total number of particles that leave $\dx/2 < x < \dx/2$ in a very small time $\dt$ is between $0$ and $N$ by some number standard deviations, $s$.

\begin{equation}
 \label{eqn:limit_condition_def}
\mu - 2s \sigma > 0 \implies N > \frac{4s^2}{\frac{D \dt}{\dx^2}}
\end{equation}

\noindent We might expect that $N$ would increase quickly with $s$, as we require smaller and smaller probability of violating conservation of mass.  However, in the continuum limit, $\dt \to 0$ and $\dx \to 0$, thus Eq. \eqref{eqn:limit_condition_def} requires that $\frac{D \dt}{\dx^2} \to 0$ regardless of how we take this limit.  Hence, Eq. \eqref{eqn:limit_condition_def} shows that we have the more stringent requirement that $N \to \infty$ as $\dt \to 0$.  Thus a continuous version of Eq. \eqref{eqn:MDE_update} is not valid for \emph{large} $N$, but strictly for $N \to \infty$.

Leaving this matter aside for the moment, we will show how one can draw an equivalence between Eq. \eqref{eqn:MDE_update} and Eq. \eqref{eqn:SDE} using Eq. \eqref{eqn:norm_approx_2}.  To obtain an expression for particle density $\rho$ rather than particle number $N$, we divide Eq. \eqref{eqn:norm_approx_2} by $\dx$, and obtain

\begin{align}
\label{eqn:norm_approx_density}
 \rho_i^{t+\dt} & =  \rho_i^t \\ \nonumber
 & + \frac{D\dt}{\dx^2}[ \rho_{i+1}^t - 2\rho_i^t + \rho_{i-1}^t] \\ \nonumber
 & + \sqrt{\frac{D\dt}{\dx^3}} \times \\ \nonumber
 & ~~~\Big[ \sqrt{\rho_{i+1}^t} \, \xi_{i+1}^t - \sqrt{\rho_i^t} \, \xi _{i,1}^t - \sqrt{\rho_i^t} \, \xi_{i,2}^t + \sqrt{\rho_{i-1}^t}  \, \xi _{i-1}^t \Big] \\ \nonumber
\end{align}

\noindent Finally, we apply the identity $\sigma_1 \, \xi  + \sigma_2 \, \xi  = \sqrt{\sigma_1^2 + \sigma_2^2} \, \xi $ to combine some of the Gaussian random variables and we are left with:

\begin{align}
\label{eqn:discrete_SDE}
 \rho_i^{t+\dt} & =  \rho_i^t \\ \nonumber
 & + \frac{D\dt}{\dx^2}[ \rho_{i+1}^t - 2\rho_i^t + \rho_{i-1}^t] \\ \nonumber
 & + \sqrt{\frac{D\dt}{\dx^3}} \Big[ \sqrt{\rho_{i+1}^t + \rho_i^t}  \, \xi _{i,1}^t - \sqrt{\rho_i^t + \rho_{i-1}^t}  \, \xi _{i,2}^t \Big] \\ \nonumber
\end{align}

Taking the continuum limit of a discrete stochastic equation such as \eqref{eqn:discrete_SDE} is not trivial.  Garcia et al. have derived a rigorous discretization of the SDE, Eq. \eqref{eqn:SDE} \cite{Garcia1987}.  We now consider an informal derivation of this same discretization.  We start with the usual discretization for the deterministic part

\begin{equation}
\label{eqn:discrete_deterministic}
 \rho_i^{t+\dt} =  \rho_i^t + \frac{D\dt}{\dx^2}[ \rho_{i+1}^t - 2\rho_i^t + \rho_{i-1}^t] 
\end{equation}

\noindent The Ito time discretization requires a factor of $\sqrt{\dt}$ for the fluctuating part

\begin{align}
\label{eqn:discrete_det_ito}
 \rho_i^{t+\dt} & =  \rho_i^t + \frac{D\dt}{\dx^2}[ \rho_{i+1}^t - 2\rho_i^t + \rho_{i-1}^t] \\ \nonumber
 & + \sqrt{dt}\frac{\partial}{\partial x}\sqrt{2D\rho} \, \xi  \\ \nonumber 
\end{align}

In the appendix, \ref{app:disc}, we show an informal way to discretize $\tfrac{\partial}{\partial x}\sqrt{\rho}\eta$.  Combining these parts, we obtain exactly Eq. \eqref{eqn:discrete_SDE}.

\section{Simulations}
The SDE, Eq. \eqref{eqn:SDE}, is derived in the thermodynamic limit, i.e. where $N$ is strictly infinite.  On the other hand, the MDE is a particle scale model.  To compare how well these two models reproduce diffusion induced fluctuations, we performed numerical simulations comparing the SDE and MDE to a particle tracking model, which we consider more fundamental and realistic.  We generated realizations of diffusion for a total time $T_\text{max}$ in a periodic domain of length $L$, initialized with $N_0$ particles distributed uniformly over the domain.  For the SDE and MDE, we discretized the domain into $N_v$ voxels of size $\dx = L/N_v$, creating an initial particle density of $n_0 = N_0/L$.  We also overlaid this grid on the particle simulation domain in order to calculate particle density.  To study the effects of time step and particle density, we varied $\dt$ and $N_0$ while fixing $D$ and $\dx$ -- which effectively defined our space and time units.

\subsection{Particle Tracking}
In the particle tracking simulations we initially filled a domain of size $L$ with $N$ particles distributed uniformly.  At each time step we used over-damped Langevin dynamics to update the positions of the particles

\begin{equation}
\label{eqn:langevin}
x_n^{t+\dt} = x_n^t + \sqrt{2 D \dt} \, \xi _n^{t + \dt} ~~~~ n = 1...N_0
\end{equation}

\noindent At each time step we counted the number of particles in each of the $N_v$ voxels defined above to obtain the particle density.

\subsection{MDE}
We evolve the MDE using Eq. \eqref{eqn:MDE_update}. Realizations of the MDE require generating multinomial random variables.  To generate multinomial random variables $n_1$ and $n_2$ from $N$ with probabilities $q_1$ and $q_2$, we used a sequential approach based on successive binomial random variables.  To generate two multinomial random variables from the multinomial distribution $\mathcal{M}[N,q_1,q_2]$, we first chose $n_1$ from from the binomial distribution $\mathcal{B}[N,q_1]$, and then chose $n_2$ from the binomial distribution $\mathcal{B}[N-n_1,q_2/(1-q_1)]$.

\subsection{SDE}
The discretization in Eq. \eqref{eqn:discrete_SDE} has a finite probability of generating negative concentrations.  Taking the absolute value of the concentration would clearly create a bias in the mean. To minimize the bias, we allowed the concentrations to be negative in the deterministic part, but took the absolute value for the square root in the fluctuations.  However, this approach would not work in simulations with reactions.  Allowing negative concentration would propagate through any reaction channel with an odd order, possibly leading to runaway negative concentrations.  Using this correction, we used the following discretization to integrate the SDE

\begin{align}
\label{eqn:discrete_SDE_2}
 \rho_i^{t+\dt} & =  \rho_i^t + \frac{D\dt}{\dx^2}[ \rho_{i+1}^t - 2\rho_i^t + \rho_{i-1}^t] \\ \nonumber
 & + \sqrt{\frac{D\dt}{\dx^2}} \Big[ \sqrt{|\rho_{i+1}^t + \rho_i^t|}  \, \xi _{i,1}^t - \sqrt{|\rho_i^t + \rho_{i-1}^t}|  \, \xi _{i,2}^t \Big]
\end{align}

\noindent where $\xi_{i,1}^t$ and $\xi_{i,1}^t$ are two different IID Gaussian random variables generated for voxel $i$ at time $t$.

\section{Results}
The statistical properties of our models are seen in the ensemble statistics.  We will use the notation $\ensav{*}{\omega}$ for ensemble averages (over realizations $\omega$) and $\ensav{*}{\Omega}$ for spatial averages (over $L$).  Let $n_\alpha(x_i)$ be the particle density at point $x_i$ in the $\alpha^{th}$ realization $\omega_\alpha$.  We define the ensemble mean particle density as $\mu(x_i) = \av{n_\alpha(x_i)}_\omega$, and ensemble particle density fluctuations as $\sigma^2(x_i) = \av{( n_\alpha(x_i) - \mu(x_i) )^2}_\omega$.  We also define a more concise number, which is the spatial domain averaged fluctuation $\bar \sigma^2 = \left< \sigma^2(x_i) \right>_\Omega$.  $\mu(x_i)$ should match the analytical steady-state solution of the deterministic diffusion equation.  However, we have no analytic expression for $\sigma^2(x_i)$.  Of our three simulation methods, the particle method is the most fundamental and realistic.  So we measure the ``convergence'' of $\mu(x_i)$ and $\bar \sigma^2$ in the MDE and SDE simulations by how well their statistics match those generated in the particle simulations.

\begin{figure}[t!]
\includegraphics[width=0.5\textwidth]{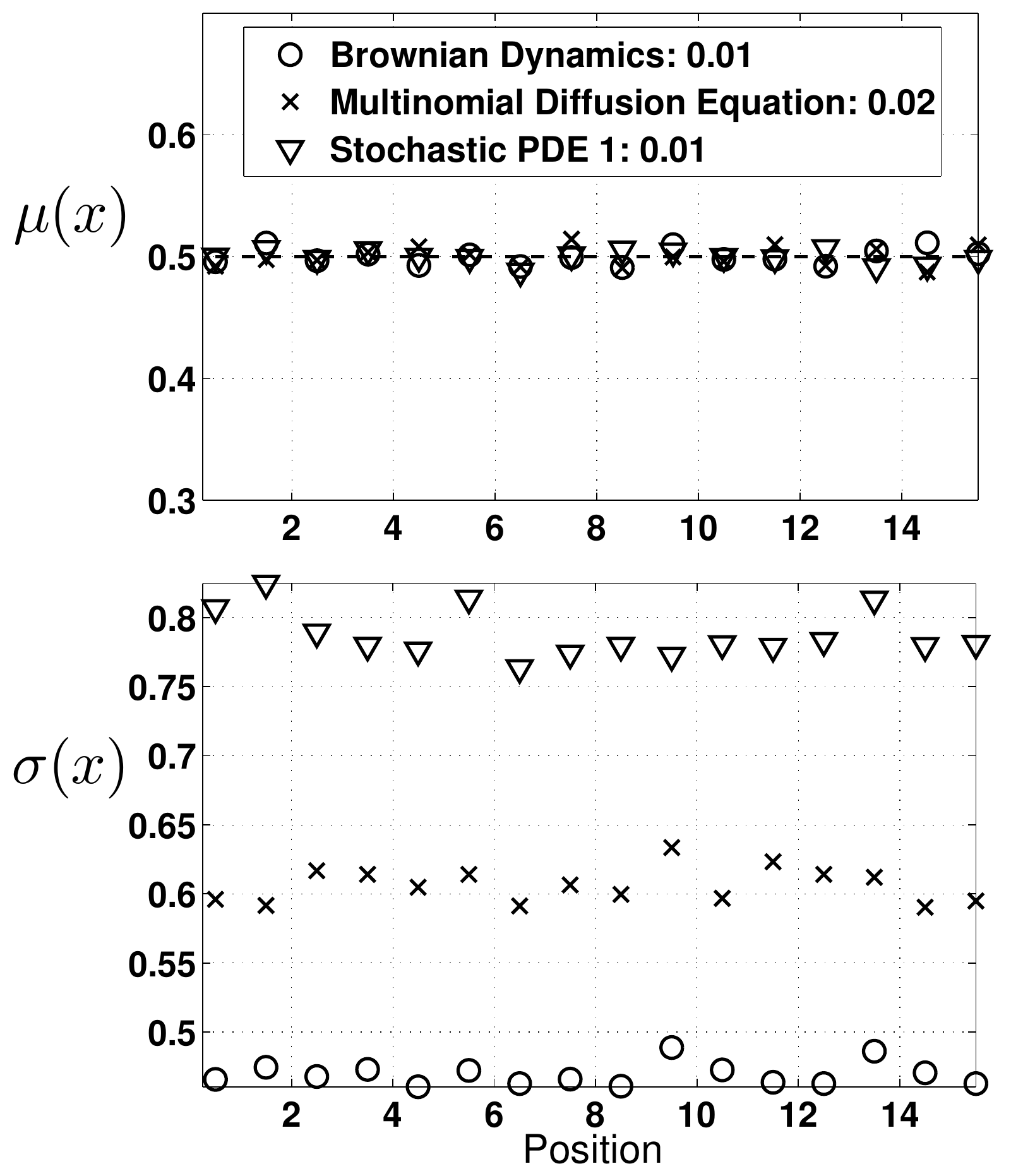}
\caption{\label{fig:ExampleData}Typical results for an ensemble (N = 8192) rangeof simulations for $D\dt=0.25$, $n_0=0.5$, $T_{max}=32$.  The legend gives a measure of fit to analytical solution (see text).}
\end{figure}

\begin{figure}[t!]
\includegraphics[width=0.5\textwidth]{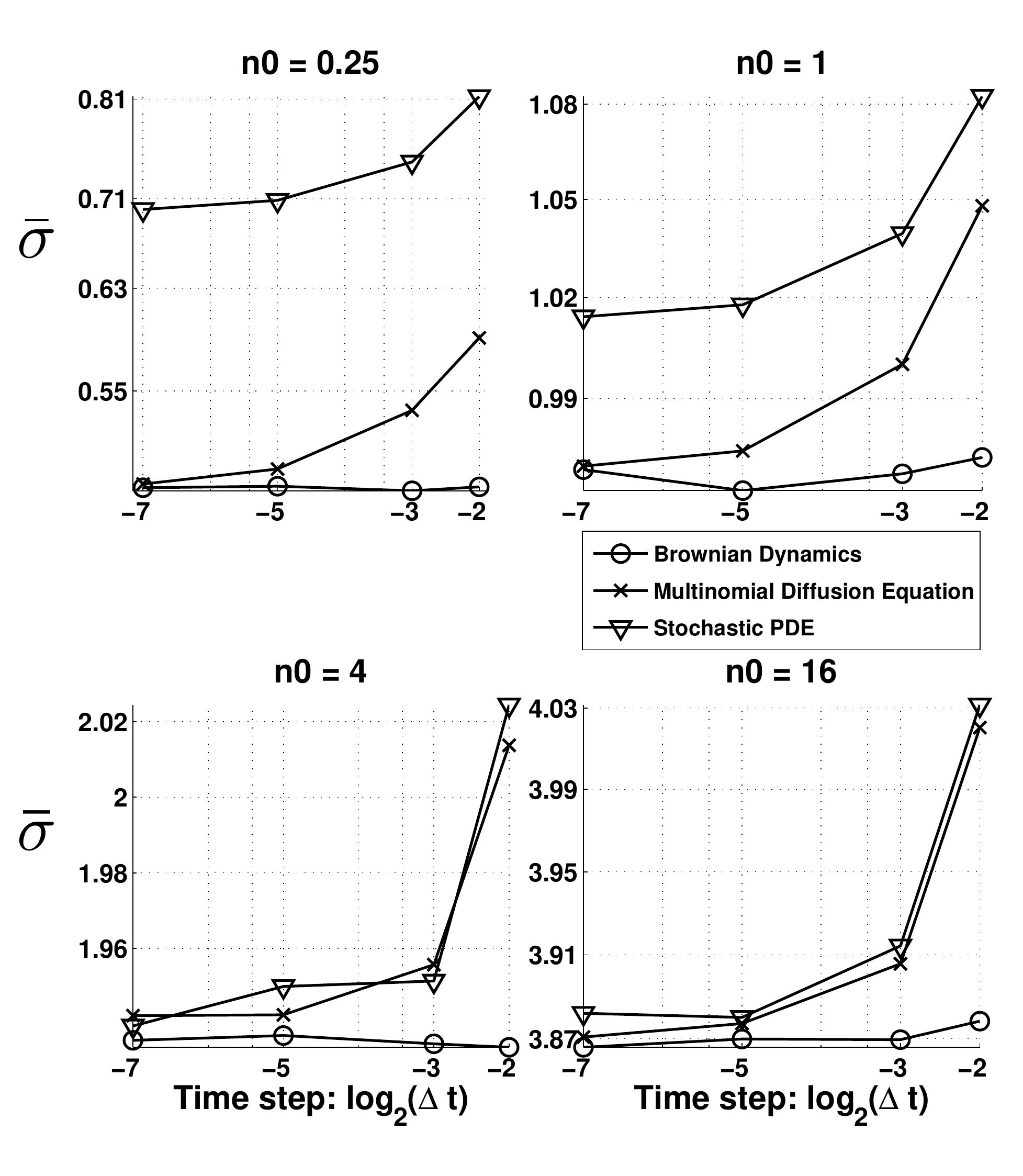}
\caption{\label{fig:ConvergenceN0}Convergence of ensemble fluctuations as a function of time step ($\dt$).  For large particle density, all models converge to the same value as the particle model as time steps decrease.  However for small particle density, the Gaussian distribution models do converge, but to the wrong value.}
\end{figure}

\begin{figure}[t!]
\includegraphics[width=0.5\textwidth]{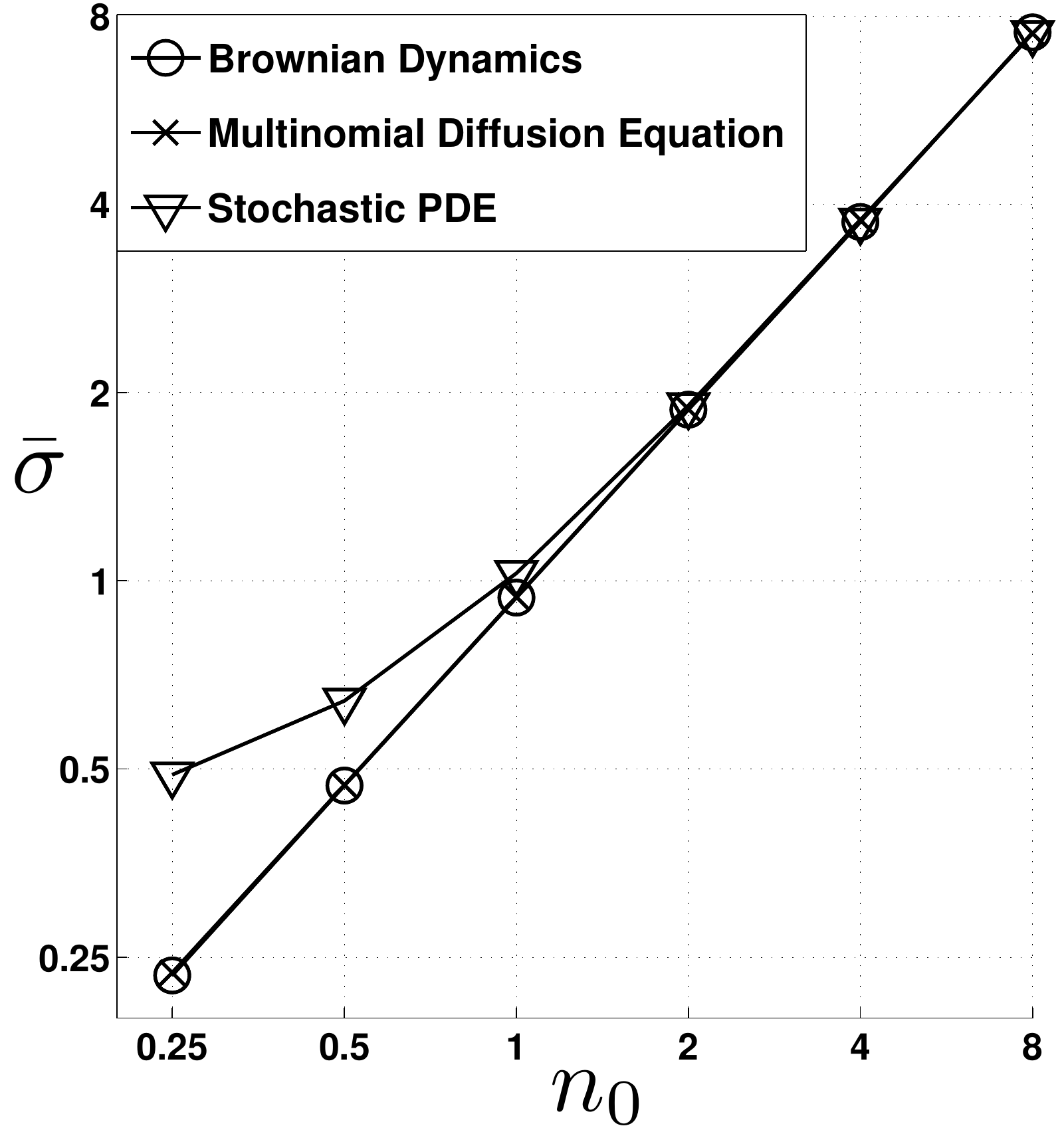}
\caption{\label{fig:SmallestDt} At the smallest value of $\dt$ we studied, all methods converge to the same ensemble fluctuations for large particle density, but diverge as smaller values.}
\end{figure}

Fig. \ref{fig:ExampleData} shows a typical result in which we see the fluctuations in the MDE simulation being somewhat larger than in the particle simulation, and the fluctuations in the SDE being larger still.  The legend box gives $\left<(\mu(x)-n_0)^2 \right>_\Omega$ as a measure of how well the the mean ensemble means agree -- a test of convergence to the analytical steady-state solution.  Fig. \ref{fig:ConvergenceN0} shows $\bar \sigma^2$ for different values of $\dt$ and $n_0$.  This plot clearly shows that $\Delta \bar \sigma^2 / \Delta (\dt) \to 0$ as $\dt \to 0$, however only the MDE converges to the same value of $\bar\sigma^2$ as the particle method.  In other words, the MDE closely replicates the ensemble statistics of a particle simulation.  On the other hand, to the extent that a numerical integration of Eq. \eqref{eqn:discrete_SDE} represents a solution of Eq. \eqref{eqn:SDE}, our data also suggests that the SDE, Eq. \eqref{eqn:SDE}, equation is not an adequate model for very small particle density.  In fig. \ref{fig:SmallestDt}, we see that the accuracy of the SDE appears to break down at about one particle/voxel.  We suspect this breakdown will occur at higher particle density in higher dimensions.

\section{Discussion}
From our simulation algorithms, we see that generating realizations of the MDE operates in a very similar way to typical finite difference methods used to solve the diffusion PDE or SDE.  This suggests possible applications of this model for numerical simulations where one needs to model some spatial regions at the particle scale and include diffusion induced fluctuations.

The grid-based MME is a somewhat more fundamental model than the MDE, and being expressed as a master equation may make it more amenable to some analytical work.  From a practical point of view, however, the MME has some limitations.  Exact sampling methods, such as the Gillespie algorithm, evolve in single particle events with with extremely small, random time steps.  Suppose a multiscale simulation has two or more disjoint regions requiring particle scale resolution.  Using the MME, each of these regions will produce independent, tiny time steps.  However, finite difference, smoothed particle hydrodynamics, finite element, and most other prevalent techniques for modeling spatiotemporal fields operate in synchronous time steps.  Not only will the MME regions become the computational bottleneck, it will be difficult to couple the independently asynchronous MME regions to synchronous regions.  Also, we previously discussed how the issue of negative concentrations causes trouble for the SDE.  The MDE, on the other hand, can integrate well with other grid-based models that operate in synchronous time steps, and requires no corrections for negative concentrations.

Recently, Alexander et al. proposed using Eq. \eqref{eqn:SDE} to couple between particle-based grid-based methods for simulating diffusion \cite{Alexander2002}.  However, if one could approximate a particle-based simulation with a much less computationally expensive grid-based method, this could be even \emph{more} valuable.  With this suggestion, we should at least minimally address the issue of performance.  For a simulation with $M$ particles, the computational expense of one time step is roughly $M$.  For a simulation with $N$ grid points, the computational expense is roughly $aN$, where $a \sim 2^d$ is some geometrical factor depending on the spatial dimension $d$ and the type of discretization.  Considering the particle density $n_0 = M/N$, we see that a grid-based simulation is less expensive when $M > aN$, i.e. $n_0 = M/N > a \ge 2$.  In this case, the critical particle density where the MDE has a performance advantage is independent of both $M$ and $N$.

In a complex system one may not know, before hand, what the particle density will be at every point in the domain at every time.  Using the MDE spares one from having to dynamically create and link particle regions to grid regions, constantly performing checks to see where and when and where to put the interfaces.  More importantly, if one includes reactions, the performance advantage increases.  With a bimolecular reaction, a particle simulation requires an additional $\sim M^2$ operations per time step; a grid-based simulation incurs only another $\sim N$ operations.  Here, a grid based simulation such as the MDE outperforms a particle simulation as long as $M^2 + M > N(a+1)$.  When many particles are involved in the simulation, $M^2 \gg M$, and (roughly) $M > \sqrt{(a+1)N}$, giving $n_0>\sqrt{(a+1)/N}$.  In this case, the critical particle density above which the MDE gains a performance advantage shrinks very steeply with the number of voxels.  For even a modest number of voxels, this quickly approaches $n_0 \ll 1$, the regime where the MDE becomes superior to the SDE in accuracy.

Monte-Carlo simulations such as the ones we have discussed are intended to generate \emph{actual realizations} of a physical model.  To insure this, we must adhere to proper space and time ordering.  This means that $E(x,x+\dx)$ in Eq. \eqref{eqn:MDE_update}, and $\sqrt{\rho^+ + \rho } \, \xi $ in Eq. \eqref{eqn:discrete_SDE} must be the \emph{same number} when used for updating the particle density at points $x$ \emph{and} $x+\dx$.  Comparing Eqs. \eqref{eqn:discrete_SDE} and \eqref{eqn:SDE} we see that proper space ordering also relates to conservation of mass: the number of particles crossing between $x$ and $x+\dx$ is the same when looked at from either side. Furthermore updating neighboring voxels with two different and independent random numbers would lead to enhanced flux fluctuations.  One often wishes to accelerate a finite difference simulation with implicit or semi-implicit methods.  It might be worthwhile to investigate if proper space and time ordering limits this.  

\section{Conclusion}
We have proposed a new discrete model of diffusion (the MDE) that sits between the multivariate master equation (MME) (a mesoscopic model) and the the classical diffusion SDE (a field model) -- a microscopic, particle model occupying an even smaller regime.  Using this model we give another, perhaps more intuitive, derivation of the classical diffusion SDE.  We perform simulations showing that at very small particle densities, the MDE very closely approximates the statistical properties of a particle simulation, while the SDE does not.  To our knowledge, this work represents the first time the classical diffusion SDE has been put to the test against a particle model.

In addition to a new theoretical model for diffusion, we suggest important practical applications of the MDE.  Although the MME is a lower-level description, it is not well suited for being coupled to other grid-based methods.  The MME is often very slow, and each domain being modeled by the MME would generate its own time step.  Particle simulations can be more efficient for modeling diffusion for small particle densities when no reactions are involved, but do not integrate seamlessly with finite-difference methods.  When even simple reactions are involved, particle methods become much less efficient, than the MDE.

\appendix 
\section{Discretization of $\tfrac{\partial}{\partial x} \sqrt{\rho}\xi$ \label{app:disc}}
Our goal is to find a numerical discretization for

\begin{equation}
 \label{eqn:derivative_of_stoch}
 \frac{\partial} {\partial x} \sqrt{\rho(x,t)} \eta(x,t)
\end{equation}

\noindent where $\eta(x)$ is unit, Gaussian, white noise.

An accepted way to form a decent discretization for the derivative of a function $f = dF/dx$ is to use a centered difference.  One can think of the centered difference discretization as estimating the slope by taking the difference of \emph{average} values of the function to the left and right of a point:

\begin{equation}
 \label{eqn:centered_difference}
 \frac{dF}{dx} \approx \frac{ \frac{F(x + \dx) + F(x)}{2} - \frac{F(x) + F(x-\dx)}{2}}{\dx}
\end{equation}

\noindent We would like to use this approach to discretize Eq. \eqref{eqn:derivative_of_stoch}.  However, due to the noisy fluctuations, Eq. \eqref{eqn:centered_difference} may not be a good enough estimate of the average.  To be more confident, we could use

\begin{align}
 \label{eqn:centered_difference_integral}
 \frac{dF}{dx} & \approx \frac{ \frac{1}{\dx}\int_x^{x+\dx} F(x')dx' - \frac{1}{\dx}\int_{x-\dx}^x F(x')dx'}{2\dx} \\ \nonumber
 & =  \frac{1}{2\dx} \int_x^{x+\dx} F(x')dx'  - \frac{1}{2\dx} \int_{x-\dx}^x F(x')dx' 
\end{align}

\noindent In our case, $F \equiv \sqrt{\rho}\eta$.  To use Eq. \eqref{eqn:centered_difference_integral}, we need to know how to integrate $\sqrt{\rho}\eta$.  For this we turn to a method due to Chandrasekhar.  Chandrasekhar approximates the integral $\int_a^b f(x) \eta(x)dx$ by

\begin{equation}
 \label{eqn:Chandra_approx}
 \int_a^b f(x) \eta(x)dx \approx  \sqrt{\dsum_{i=1}^N f^2(x_i) \dx} ~ \eta(x)
\end{equation}

\noindent where $\dx = \frac{b-a}{N}$ and $x_i = a + \dx(i-1)$.  We now approximate the first integral in Eq. \eqref{eqn:centered_difference_integral} by setting $N=2$, $a = x$, and $b = x+\dx$, and the second integral by setting $a = x-\dx$, and $b = x$. In principle, it makes sense to index $\eta$ at any point between $x$ and $x+\dx$.  In this way, we finally obtain 
%

\begin{multline}
 \label{eqn:disc_derivative}
\frac{\partial}{\partial x} \sqrt{\rho(x,t)}\eta(x,t) \approx \\
\tfrac{ \sqrt{\rho(x+\dx,t) + \rho(x,t)} \, \eta(x+\dx,t) - \sqrt{\rho(x,t) + \rho(x-\dx,t}\, \eta(x-\dx,t)}
{2\sqrt{\dx^3}}
\end{multline}

\bibliographystyle{plain}
\bibliography{StochFD}

\begin{thebibliography}{10}

\bibitem{Note1}
Just as a single Gaussian approximates a binomial.

\bibitem{Note2}
Interestingly, we see that the Courant condition \cite {press_numerical_1992}
  for the stability of numerical discretization equivalent to the deterministic
  part of the Gaussian version corresponds to conservation of mass in the
  model.

\bibitem{Alexander2002}
F~J Alexander, A~L Garcia, and D~M Tartakovsky.
\newblock Algorithm refinement for stochastic partial differential equations:
  I. linear diffusion.
\newblock {\em Journal of Computational Physics}, 182(1):47 -- 66, 2002.

\bibitem{Awazu2010}
A~Awazu and K~Kaneko.
\newblock Discreteness-induced slow relaxation in reversible catalytic reaction
  networks.
\newblock {\em Phys. Rev. E}, 81(5):051920, May 2010.

\bibitem{Breiman1969}
Leo Brieman.
\newblock {\em Probability and Stochastic processes with a veiw towards
  applications}.
\newblock Houghton Mifflin Company, 1969.

\bibitem{Dewit2007}
A~DeWit.
\newblock {\em Spatial Patterns and Spatiotemporal Dynamics in Chemical
  Systems}.
\newblock John Wiley \& Sons, Inc., 2007.

\bibitem{Garcia1987}
A~L Garcia, M~Mansour, G~C Lie, and E~Cementi.
\newblock Numerical integration of the fluctuating hydrodynamic equations.
\newblock {\em Journal of Statistical Physics}, 47:209--228, 1987.
\newblock 10.1007/BF01009043.

\bibitem{Gardiner1996}
C~W Gardiner.
\newblock {\em Handbook of Stochastic Methods: For Physics, Chemistry and the
  Natural Sciences (Springer Series in Synergetics)}.
\newblock Springer, November 1996.

\bibitem{Gillespie2007}
Daniel~T Gillespie.
\newblock Stochastic simulation of chemical kinetics.
\newblock {\em Annu Rev Phys Chem}, 58:35--55, 2007.

\bibitem{Gopich2001}
I~V Gopich, A~A Ovchinnikov, and A~Szabo.
\newblock Long-time tails in the kinetics of reversible bimolecular reactions.
\newblock {\em Phys. Rev. Lett.}, 86(5):922--925, Jan 2001.

\bibitem{Hecht2010}
Inbal Hecht, David~A. Kessler, and Herbert Levine.
\newblock Transient localized patterns in noise-driven reaction-diffusion
  systems.
\newblock {\em Phys. Rev. Lett.}, 104(15):158301, Apr 2010.

\bibitem{Keizer1987}
Joel Keizer.
\newblock {\em {Statistical Thermodynamics of Nonequilibrium Processes}}.
\newblock Springer, 1 edition, July 1987.

\bibitem{VanKampen2007}
N.~G. Van~Kampen.
\newblock {\em {Stochastic Processes in Physics and Chemistry, Third Edition
  (North-Holland Personal Library)}}.
\newblock North Holland, 3 edition, May 2007.

\bibitem{Togashi2001}
Yuichi Y~Togashi and K~Kaneko.
\newblock Transitions induced by the discreteness of molecules in a small
  autocatalytic system.
\newblock {\em Phys. Rev. Lett.}, 86(11):2459--2462, Mar 2001.

\end{thebibliography}

\end{document}